\documentclass[fleqn,twoside]{article}
\usepackage{espcrc2}

\usepackage{graphicx}
\usepackage[figuresright]{rotating}

\usepackage{amssymb}

\newcommand{\AmS}{{\protect\the\textfont2
  A\kern-.1667em\lower.5ex\hbox{M}\kern-.125emS}}

\title{Search for neutrinoless double beta decay with the NEMO-3 detector: first results}

\author{X. Sarazin\address{Laboratoire de l'Acc\'elerateur Lin\'eaire, 
        IN2P3-CNRS et Universit\'e de Paris Sud, B\^at. 200 - BP 34 \\ 
        91898 Orsay Cedex, France} %
         on behalf of the NEMO collaboration}

\begin{document}
\thispagestyle{empty}

\begin{abstract}
The NEMO-3 detector, which has been operating in the Fr\'ejus Underground 
Laboratory since 
February 2003, is devoted to  searching for neutrinoless double beta 
decay ($\beta\beta$0$\nu$). 
The expected performance of the detector has been successfully achieved.
Half-lives of the two neutrinos double beta decay ($\beta\beta 2 \nu$) 
have been measured for $^{100}$Mo, $^{82}$Se, $^{96}$Zr, $^{116}$Cd 
and $^{150}$Nd.
After 265 days of data collection from February 2003 until March 2004, 
no evidence for neutrinoless double beta decay ($\beta\beta 0 \nu$) 
was found from  $\sim$7~kg of $^{100}$Mo and $\sim$1~kg of $^{82}$Se. 
The corresponding lower limits for the half-lives are $3.5 \times 10^{23}$ 
years at 90\% C.L for $^{100}$Mo and $1.9 \times 10^{23}$ years for  $^{82}$Se.
Limits for the effective Majorana neutrino mass are $<$\hspace{-0.5mm}$m_{\nu}$\hspace{-0.5mm}$> < 0.7-1.2$~eV for $^{100}$Mo and \linebreak $<$\hspace{-0.5mm}$m_{\nu}$\hspace{-0.5mm}$> < 1.3-3.2$~eV for $^{82}$Se.
Radon is the dominant background today and a Radon-free purification system will be in operation by the end of september 2004. 
The NEMO-3 expected sensitivity after 5 years of data  is 0.2 eV. \\

\end{abstract}

\maketitle

\section{Introduction}

Neutrinoless double beta decay ($\beta\beta$0$\nu$) is a process beyond the 
Standard Model which violates lepton number by 2 units. The discovery of 
this decay would be experimental proof that the neutrino is a Majorana 
particle. It would also constrain the mass spectrum and the absolute mass
of the neutrinos. The NEMO-3 detector is devoted to searching 
for $\beta\beta 0 \nu$ decay with the direct detection of the two 
electrons from $\beta\beta$ decay by a combination of a tracking 
device and a calorimeter. 

\section{The NEMO-3 detector}

The NEMO-3 detector \cite{Augier2004}, installed in the Fr\'ejus 
Underground Laboratory (LSM, France) is a cylinder divided into 20 
equal sectors. 
The isotopes present inside the detector in the form of very thin foils (30-60 $mg/cm^2$) are $^{100}Mo$ (6914~g), 
$^{82}Se$ (932~g), $^{116}Cd$ (405~g), $^{130}Te$ (454~g), 
natural Te (491~g), $^{150}Nd$ (34~g), $^{96}Zr$ (9~g), 
$^{48}Ca$ (7~g) and Cu (621~g). Natural Te and Cu are devoted to measuring 
the external background. The sources have been purified to reduce their content of $^{214}$Bi and $^{208}$Tl.
On both sides of the sources, there is a gaseous tracking detector. 
It consists of $6180$ open drift cells operating in the Geiger mode 
regime (geiger cells) which gives three-dimensional track reconstruction. 
To minimize the multiple scattering, the gas is a mixture of 95\% He, 4\% 
ethyl alcohol and 1\% Argon. 
The wire chamber is surrounded by a calorimeter which consists of 
$1946$ plastic scintillator blocks coupled to very low radioactive 
photomultipliers (PMTs) especially developed by Hamamatsu. 
A solenoid surrounding the detector produces a 25 G magnetic field 
in order to recognize electrons and distinguish them from positrons. 
Finally an external shield of 18 cm of low radioactivity iron covers the 
detector to reduce external $\gamma$ and thermal neutrons. 
Outside the iron, a water shield and a wood shield  thermalize neutrons.
Thus the combination of a tracking detector, a calorimeter and a magnetic 
field allows the identification of electrons, positrons, $\gamma$ 
and $\alpha$ particles. 

\section{Performance of the detector}

Within the tracking detector, 99.5\% of the geiger cells are functioning 
normally. The vertex resolution in the 2 electron channel has been measured 
using the simultaneous 2 electron conversion of $^{207}Bi$ sources 
placed inside the detector. 
The  resolution on the distance between 
the two reconstructed tracks is $\sigma_{t} = 0.6$ cm in the transverse plane 
and $\sigma_{l} = 1.3$ cm in the longitudinal plane. The ambiguity 
between $e^-$ and $e^+$ is 3\% at 1 MeV. 

Within the calorimeter, 97\% of the PMTs coupled to scintillators 
are functioning correctly. 
The energy resolution, measured every $\sim$40 days using $^{207}Bi$ 
sources, is 15\% (FWHM) at 1 MeV for the 5'' 
PMTs on the external wall and 17\% for the 3'' PMTs on the internal wall. A daily laser 
survey controls the gain stability of each PMT. The efficiency to 
detect a $\gamma$ at 500 keV is about 50\% with a threshold of 30 keV.
The time resolution measured with the 2 electron channel, 
is 250 ps at 1 MeV which is much smaller than the time-of-flight of a 
crossing electron that is larger than 3 ns. Thus external crossing 
electrons are totally rejected. 

In conclusion, the expected performance of the NEMO-3 detector has 
been successfully achieved.

\section{Measurement of $\beta\beta 2 \nu$ decays with several nuclei}

\begin{figure}[ht]  
\includegraphics[scale=0.31]{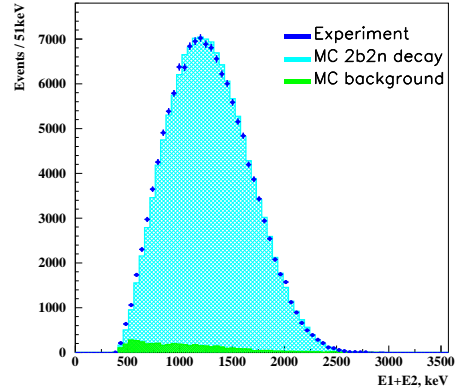} 
\vspace{-1cm}
\caption{\label{esum} Spectrum of the energy sum of the two electrons from $6.914$~kg of $^{100}$Mo after 241 days of data collection.}
\end{figure}

The detector has been running since February 2003. The trigger 
configuration requires at least 1 PMT with an energy above 150 keV and 3 
active geiger cells. The trigger rate is $\sim$7~Hz. A $\beta\beta$ event 
is an event with 2 tracks coming from the same 
vertex on the foil. Each track  is associated to a fired scintillator with 
a good internal time-of-flight hypothesis, and the curvature corresponds 
to a negative charge. Such a $\beta\beta$ event is detected 
every $\sim$1.5 minutes. 
After 241 days of data analysed, more than 140,000 $\beta\beta 2 \nu$ 
events from $\sim$7 kg of $^{100}$Mo  have been measured. 
Figure~\ref{esum} shows the spectrum of the summed energy of the two 
electrons for $^{100}$Mo after background subtraction which is in agreement 
with the expected spectrum from $\beta\beta 2 \nu$ simulation. 


The subtracted background is very low, corresponding to a very high 
signal-to-background ratio of 46. The preliminary value of the 
measured half-life is $7.72 \pm 0.02(stat) \pm 0.54(syst) 10^{18}$ y. 
The $\beta\beta 2 \nu$ decay has also been measured for $^{82}$Se, 
$^{96}$Zr, $^{116}$Cd and $^{150}$Nd. Preliminary results are 
summarized in table~\ref{table_bb2nu}.


\begin{table*}[htb]
\caption{Preliminary results of the measurements of $\beta\beta 2 \nu$ decays}
\label{table_bb2nu}
\newcommand{\m}{\hphantom{$-$}}
\newcommand{\cc}[1]{\multicolumn{1}{c}{#1}}
\renewcommand{\tabcolsep}{1pc} 
\renewcommand{\arraystretch}{1.2} 
\begin{tabular}{@{}llllll}
\hline
Isotope       & mass & days    & number of           & Signal/ & $T_{1/2}(\beta\beta 2 \nu)$  \\
              & (g)  & of data & $\beta\beta$ events & Background    & (years) \\
\hline
$^{82}Se$     & 932  & 241.5   & 2385   & 3.3  & $10.3 \pm 0.2(stat)  \pm 1.0(syst)  10^{19}$ y.  \\
$^{96}Zr$     & 9.4  & 168.4   & 72     & 0.9  & $2.0  \pm 0.3(stat)  \pm 0.2(syst)  10^{19}$ y.  \\
$^{100}Mo$    & 6914 & 241.5   & 145245 & 45.8 & $7.72 \pm 0.02(stat) \pm 0.54(syst) 10^{18}$ y.  \\
$^{116}Cd$    & 405  & 168.4   & 1371   & 7.5  & $2.8  \pm 0.1(stat)  \pm 0.3(syst)  10^{19}$ y.  \\
$^{150}Nd$    & 37.0 & 168.4   & 449    & 2.8  & $9.7  \pm 0.7(stat)  \pm 1.0(syst)  10^{18}$ y.  \\
\hline
\end{tabular}\\[2pt]
\end{table*}

\section{Study of the background in the $\beta\beta 0 \nu$ energy window}

After almost 1 year of data, the level of each component of the 
background has been directly measured using different channels in the data. 

External $^{214}$Bi and $^{208}$Tl backgrounds (mostly inside the PMTs) 
have been measured by searching for external ($e^-$,$\gamma$) events in the 
data. The total reconstructed activity of  $^{208}$Tl is $\sim$40 Bq and is in 
agreement with previous HPGe measurements of a sample of the PMTs glass. 
The expected number of $\beta\beta 0 \nu$-like events is negligible, 
$\lesssim10^{-3}$ counts kg$ ^{-1}$y$^{-1}$ in the $[2.8-3.2]$ MeV 
energy window where $\beta\beta 0 \nu$ signal is expected.

External neutrons and high energy $\gamma$ backgrounds have been 
measured by searching for internal ($e^-$,$e^-$) events above 4 MeV. 
Only 2 events have been observed in 265 days of data collection, as expected 
in the Monte-Carlo.\newpage
\noindent
 This background is also negligible, 
$\lesssim0.02$ counts kg$ ^{-1}$y$^{-1}$ in the $[2.8-3.2]$ 
MeV $\beta\beta 0 \nu$ energy window.

The level of $^{208}$Tl impurities inside the molybdenum sources has 
been measured by searching for internal ($e^-$,$\gamma$) and 
($e^-$,$\gamma\gamma$) events. An activity of $\sim100 \mu$Bq/kg has been 
measured in good agreement with the previous Ge measurements done 
before installing the sources in the detector. This corresponds to 
an expected number of $\beta\beta 0 \nu$-like events of $\sim 0.1$ counts 
kg$ ^{-1}$y$^{-1}$ in the $[2.8-3.2]$ MeV $\beta\beta 0 \nu$ energy window.

The expected level of background due to the tail of the 
$\beta\beta 2 \nu$ is $\sim0.3$ counts kg$ ^{-1}$y$^{-1}$ in 
the $[2.8-3.2]$ MeV $\beta\beta 0 \nu$ energy window.

The dominant background today is the radon inside the tracking chamber 
due to a low level of diffusion of the radon inside the laboratory 
($\sim$15~Bq/m$^3$) into the detector. Two independant measurements have 
been carried out. 
A high efficiency radon detector has measured radon in the NEMO-3 gas. 
Radon can also be measured directly by searching for ($e^-$,$\alpha$) events 
in the NEMO-3 data. Indeed the tracking detector allows the detection of the
delayed tracks (up to 700~$\mu$s) in order to tag delayed alphas emitted 
by  $^{214}$Po in the $Bi-Po$ process. Both measurements are in good 
agreement and indicate a level of radon inside the detector of $\sim$20-30~mBq/m$^3$. 
This radon contamination corresponds to an expected number of 
$\beta\beta 0 \nu$-like events of $\sim$1 count kg$ ^{-1}$y$^{-1}$ 
in the $[2.8-3.2]$ MeV $\beta\beta 0 \nu$ energy window, a factor $\sim$10 too 
high to reach the expected sensitivity.  A Radon-free purification system, 
designed to reduce radon contamination by a factor $\sim$100 will be 
in operation by the end of September 2004.

\section{Preliminary results on the limit of $\beta\beta 0 \nu$ decay with $^{100}$Mo and $^{82}$Se}

\begin{figure}[ht]  
\includegraphics[scale=0.35]{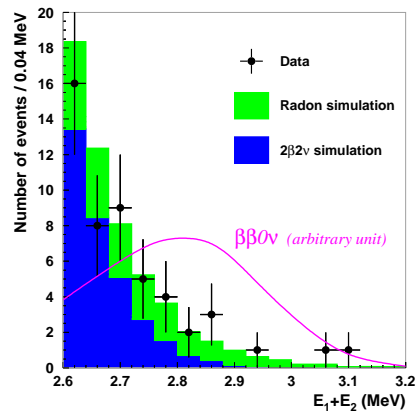} 
\vspace{-1cm}
\caption{\label{esum_mo100} Spectrum of the energy sum of the two electrons above 2.6 MeV from $6.914$~kg of $^{100}$Mo after 265 days of data collection.}
\end{figure}

\begin{figure}[ht]  
\includegraphics[scale=0.35]{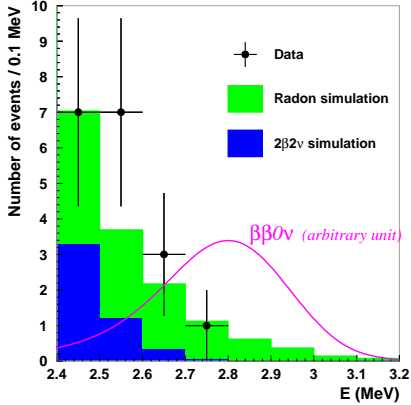} 
\vspace{-1cm}
\caption{\label{esum_se82} Spectrum of the energy sum of the two electrons above 2.4 MeV from $0.932$~kg of $^{82}$Se after 265 days of data collection.}
\end{figure}

Figures~\ref{esum_mo100} and ~\ref{esum_se82} show the 
spectrum of the energy sum of the two electrons in the $\beta\beta 0 \nu$ energy window after 
265 days of data collection with  $6.914$~kg of $^{100}$Mo and
 $0.932$~kg of $^{82}$Se respectively.

The number of two electron events observed in the data is in agreement 
with the expected number of events from $\beta\beta 2 \nu$  
and the radon simulations. 

\begin{figure}[ht]  
\includegraphics[scale=0.35]{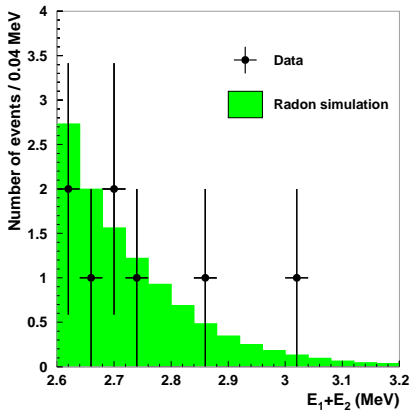} 
\vspace{-1cm}
\caption{\label{esum_cute} Spectrum of the energy sum of the two electrons above 2.6 MeV from copper and tellerium foils after 265 days of data collection.}
\end{figure}

In the energy window $[2.8-3.2]$ MeV, the expected background is 
$7.0\pm1.7$ and 8 events have been observed from $^{100}$Mo.
To check independently the dominant radon contribution above 2.8 MeV, the same 
energy sum spectrum (Fig.~\ref{esum_cute}) has been plotted for the 
two electrons emitted from the copper and tellurium foils where no 
background except radon is excepted: the data are in agreement with 
the radon simulation.


The NEMO-3 detector is able to measure not only the energy sum ($E_{tot}$) of the 
2$e^-$ events but also the single energy ($E_{min}$ energy of the $e^-$ of minimum energy) 
and the angle between the two tracks ($cos\theta$). Moreover the level 
of each component of background can be measured through studies of
different channels, as explained above. Therefore a maximum likelihood 
analysis has been applied on 2$e^-$ events above 2 MeV using these three 
variables \cite{Etienvre2003}. A three-dimensional probability distribution 
function, $P^{3D}$, can be written as:\\
\hspace*{3mm}$P^{3D}\, = \, P(E_{tot})\,P(E_{min}/E_{tot})\,P(cos\theta /E_{min})$\\
where $P(E_{min}/E_{tot})$ and $P(cos\theta/E_{min})$ are two 
conditional probability distribution functions. The likelihood is defined as\\
\hspace*{1cm}${\cal L} = \,  \prod_{i\, =\,1}^{N_{tot}} (\sum_{k=1}^8 \, x_k P^k_{3D})$\\
where $k$ corresponds to one of the 8 contributions: $\beta\beta 0 \nu$, $\beta\beta 2 \nu$, Radon, external and internal $^{214}$Bi and $^{208}$Tl, and neutrons. 
$x_k$ is the ratio of the number of 2$e^-$ events due to  the process $k$ to 
the total number of observed events $N_{tot}$, $P^k_{3D}$ is built using simulated events of the contribution $k$. The only free parameter is $x_{0\nu}$. 

With 265 days of data, limits obtained with the likelihood analysis 
are $T_{1/2}(0\nu)>3.5 \times 10^{23}$ years at 90\% C.L for $^{100}$Mo 
and $1.9 \times 10^{23}$ years for  $^{82}$Se. The corresponding upper 
limits for the effective Majorana neutrino mass range from 0.7 to 1.2 eV 
for $^{100}$Mo and 1.3 to 3.6 eV for $^{82}$Se depending on the nuclear matrix elements 
\cite{Simkovic1999,Stoica2001,Rodin2003,Caurier1996}.
Limit on Majoron is $T_{1/2}(M)>1.4 \times 10^{22}$ years at 90\% C.L, corresponding to a limit of $\chi < (5.3-8.5) \times 10^{-5}$ \cite{Simkovic1999,Stoica2001}.

\section{Conclusions}

The NEMO-3 detector has been running reliably since February 2003. 
The $\beta\beta 2 \nu$ decay has been measured for $^{82}$Se, $^{96}$Zr, 
$^{100}$Mo, $^{116}$Cd and $^{150}$Nd. 
All components of the background in the $\beta\beta 0 \nu$ energy window 
have been measured directly using different channels in the data. The 
energy sum spectrum of $2e^-$ events is in agreement with the 
simulations.
After 265 days of data, no evidence for $\beta\beta 0 \nu$ decay is found from the $6.914$~kg 
of $^{100}$Mo and $0.932$~kg of $^{82}$Se. A likelihood analysis 
gives an upper limit for the effective  neutrino 
mass $<$\hspace{-0.5mm}$m_{\nu}$\hspace{-0.5mm}$> < 0.7-1.2$~eV 
for $^{100}$Mo.
Radon is the dominant background today. A Radon-free purification 
system will be in operation by the end of September 2004. 
After radon purification and 5 years of data collection, the expected sensitivity will be 
$T_{1/2}(0\nu) > 4 \times 10^{24}$ years at 90\% C.L 
for $^{100}$Mo and $8 \times 10^{23}$ years for $^{82}$Se, 
corresponding to $<$\hspace{-0.5mm}$m_{\nu}$\hspace{-0.5mm}$> < 0.2-0.35$~eV 
for $^{100}$Mo and $<$\hspace{-0.5mm}$m_{\nu}$\hspace{-0.5mm}$> < 0.65-1.8$~eV 
for $^{82}$Se.



\end{document}